\documentclass[a4paper,11pt]{article}
\pdfoutput=1 
\usepackage{jinstpub} 
\usepackage{multirow} 
\usepackage{makecell} 
\usepackage{textcomp}
\usepackage{graphicx}
\usepackage{epstopdf}
\usepackage{overpic}
\usepackage{url}
\title{Measurement of the relative Quantum Efficiency of Hamamatsu model R5912-20MOD photomultiplier tubes at liquid argon temperature}

\author[a,b]{Q. Zhao,}
\author[a,1]{M.Y. Guan,%
\note{Corresponding author.}}
\author[a,b]{P. Zhang,}
\author[a,b]{C.G. Yang,}
\author[a,b,c]{J.J. Li,}
\author[a,b]{Y.T. Wei,}
\author[a]{L. Wang,}
\author[a,b]{Y.Y. Gan}
\author[a,b]{and W.X. Xiong}

\affiliation[a]{Institute of High Energy Physics, Chinese Academy of Sciences, Beijing, 100049, China}
\affiliation[b]{University of Chinese Academy of Sciences, Beijing, 100049, China}
\affiliation[c]{North China Electric Power University, Beijing, 100096, China}

\emailAdd{dreamy\_guan@ihep.ac.cn}

\abstract{The model R5912-20MOD photomultiplier tube(PMT) is made for cryogenic application by Hamamatsu. In this paper, we report on the measurement of relative quantum efficiency (QE) of this model PMT at liquid argon(LAr) temperature. Furthermore, a specially designed setup and relevant test method are introduced. The relative QE is measured in visible wavelengths with the PMT emerged in high purity nitrogen atmosphere. The results show that the change of QE at LAr temperature is within about 5$\%$ compared with room temperature around 420 nm. However, the QE increases about 10$\%$ in the shorter wavelength range and decreases significantly after 550 nm.}


\begin{document}
\maketitle
\flushbottom

\section{Introduction}

The photomultiplier tube(PMT) is the most commonly used photoelectric device in particle physics experiments. Many experiments are using PMTs, including JUNO\cite{JUNO,JUNO-zhang}, DEAP-3600\cite{DEAP}, XMASS\cite{XMASS}, PandaX\cite{PandaX}, XENON1T\cite{R11410-21} and so on. Due to different application environments and different spectral response range, there are various types of PMTs. For the liquid noble gas detector, the working temperature is much lower than the room temperature. Some key characteristics of the low temperature operation PMTs given by the manufacturer usually are tested at room temperature, therefore, it is necessary to make specific test for the proper application temperature. One of the key characteristic parameters is the quantum efficiency(QE). The QE of PMT R11410-10\cite{R11410-10} and R11410-21\cite{R11410-21} for liquid xenon experiments were measured at liquid xenon temperature. The results show that during the PMT cooldown from room temperature to -110\textcelsius (liquid xenon temperature) the QE increases by a factor of 1.1-1.15 at 175 nm and the QE variation with temperature is wavelength dependant\cite{R11410-10}. However, for the model R5912 series PMT, the literature is mainly focused on R5912\cite{R5912}, R5912-MOD\cite{R5912-MOD-01,R5912-MOD-02} and R5912-02MOD\cite{R5912-02MOD-01,R5912-02MOD-02}. The model R5912-20MOD PMT is made for working at liquid argon(LAr) temperature by Hamamatsu. According to the datasheet, at 25\textcelsius, the QE is between 15$\%$ and 17$\%$ at 420 nm. The QE of R5912-20MOD PMT at LAr temperature will change, comparing with liquid xenon PMTs. The DUNE experimental group has tested the model R5912-20MOD PMT at liquid nitrogen temperature, but the performance parameters of the test don't involve QE\cite{R5912-20}. Therefore, it is necessary to quantify the relative change of QE at LAr temperature compared with room temperature. In this paper, a set of dedicated devices is introduced for measuring the relative QE. And, two PMTs are randomly selected for relative QE test. The results show that the relative changes of QE are less than 5$\%$ at LAr temperature compared with room temperature at 420 nm.

According to traditional QE measurement methods, it needs a reference tube with known QE\cite{R11410-10,R5912,JUNO-zhang,ZHONG}. Two beam of light from a light source with fixed intensity ratio is used to illuminate the reference tube and the test tube respectively. The QE is obtained according to the ratio of each photocathode current\cite{ZHONG}. In the literature\cite{R11410-10}, parts of the test PMT except photocathode are covered with copper and cooled by heat conduction. There is also a copper coil placed in the container by flowing liquid nitrogen through the coil to cool the system\cite{R11410-21}. To measure the PMT QE varying with different wavelengths, the monochromator is used to get different wavelengths lighted by the tungsten lamp and the deuterium lamp. In this paper, a set of surface-mounted LEDs with different spectral ranges are used as light source, and a linear optical photosensor is used to monitor the light intensity for the reference as well. We mainly focuses on the measurement of relative QE. As for the absolute QE, it is the direction of improvement in future experiments.

\section{Test method and experimental setup}

By definition, QE is the ratio of the number of photoelectrons emitted from the photocathode and the number of incident photons\cite{PMT}. Given the continuous light source, the photoelectrons emitted from the photocathode is proportional to the measured photocathode current. The incident photons is determined by the light intensity illuminating the PMT. Therefore, the measurement of photocathode current of PMT and the measurement of light intensity are the key issues for the relative QE test system.

\begin{figure}[htbp]
  \centering
  \includegraphics[width=9cm]{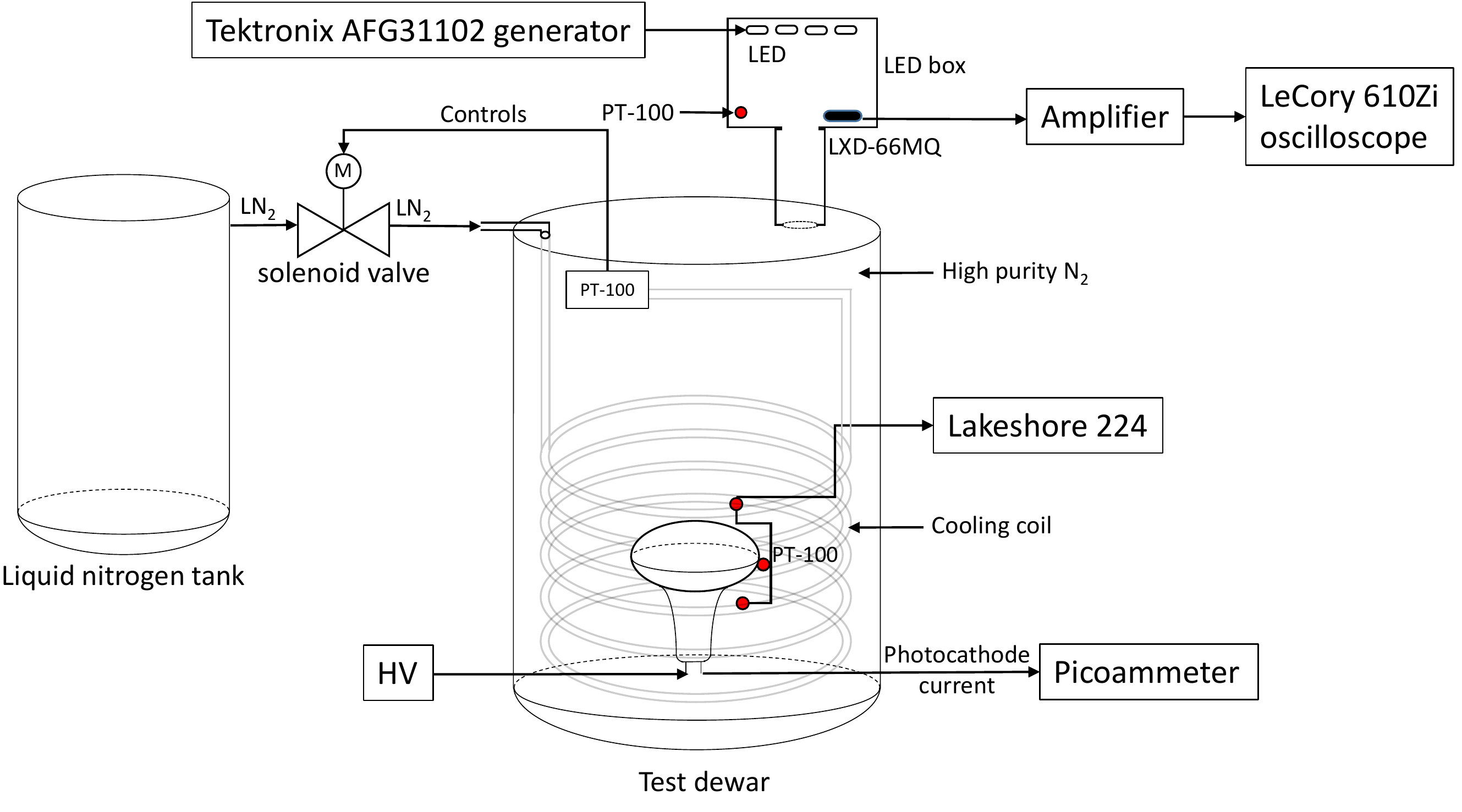}
  \caption{\label{fig:test_system} The schematic diagram of relative QE test system.}
\end{figure}

To test relative QE at LAr temperature, a test system as shown in Figure.$\ref{fig:test_system}$ is set up. A PT-100 temperature sensor is used to monitor the temperature of the outlet of cooling coil. The liquid nitrogen flow is controlled by a PID controller and a solenoid valve to provide precise temperature regulation for the test dewar. The test PMT is placed on the support structure in the gastight dewar which is filled with high purity nitrogen with 1 atm pressure. As the light source tested is in the visible light band, the nitrogen is fully transparent according to the description in\cite{N2}. In addition, Rueger's formula\cite{refractivity} is authoritative for the description of atmospheric refractive index. According to the formula, the refractive index of nitrogen changes in the order of ppm magnitude when the temperature changes from room temperature to LAr temperature. So the light intensity distribution in the test dewar remains unchanged at different temperatures. Three PT-100s are fixed around the test PMT to monitor the temperature: one is set about 5 cm above the top of the PMT, one touches the PMT glass shell at the maximum diameter, and the other is about 2cm next to the junction of the head and tail of the PMT. The Lakeshore-224 temperature monitor is used to record the temperature data. A Keithley 6485 picoammeter is used to measure the photocathode current with all the PMT dynodes being short-circuited. 

\begin{table}
    \centering
    \caption{\label{LED} The parameter table of LED.}
    \begin{tabular}{ccccccc}
    \hline
        \makecell[c]{LED No.} & \makecell[c]{Wavelength(nm)} & \makecell[c]{Rated voltage(V)} & \makecell[c]{Luminance(lm)} & \\ \hline
        1 & 360 & 3.4-3.6 & 2-5 & \\
        2 & 370 & 3.3-3.5 & 3-4 &\\
        3 & 380-385 & 3.3-3.5 & 4-5 &\\
        4 & 385-390 & 3.3-3.5 & 5-6 &\\
        5 & 400-405 & 3.3-3.4 & 8-10 &\\
        6 & 405-410 & 3.3-3.5 & 10-12 &\\
        7 & 410-415 & 3.3-3.5 & 10-12 &\\
        8 & 425-430 & 3.3-3.5 & 10-12 &\\
        9 & 440-450 & 3.2-3.4 & 50-60 &\\
        10 & 460-470 & 3.2-3.4 & 50-60 &\\
        11 & 500-510 & 3.0-3.1 & 130-140 &\\
        12 & 520-525 & 3.2-3.4 & 160-180 &\\
        13 & 590-592 & 2.2-2.4 & 90-100 &\\
        14 & 595-610 & 2.2-2.4 & 90-100 &\\
        15 & 620-625 & 2.2-2.4 & 100-110 &\\ \hline
    \end{tabular}
\end{table}
 15 LEDs are welded onto an aluminum circuit board to form an array, which is fixed on one end of an aluminum alloy box (LED box). The basic parameters of 15 LEDs with different spectral ranges are shown in Table.$\ref{LED}$. Each LED is driven separately by the Tektronix AFG31102 dual-channel arbitrary function generator through a DIP switch. During the test, the working mode of the function generator is set to the square wave mode with a frequency of 1kHz and a duty cycle of 50$\%$\cite{JUNO-zhang,R11410-21}. A 3cm diameter hole is opened on the other end of the LED box for the output of light. A piece of Tyvek film is adopted to attached on the inner wall of the LED box to condense light. The light enters the test dewar through a view port with quartz glass to illuminate the photocathode surface of PMT. In the LED box, a PT-100 temperature sensor is placed to monitor temperature changes. Since the LED box is place about 30cm away from the top flange of the test dewar, the temperature shows nearly no change during the whole test. A model LXD-66MQ linear optical photosensor\cite{Optical_PS,SiPS} is placed on the side of the hole in the LED box to monitor the change of LED light intensity. The signal of optical photosensor is converted by the I-V conversion amplifier and recorded by the oscilloscope. The equivalent light intensity data is derived from the average peak-to-peak value of the output signal waveform of the optical photosensor.

\begin{figure}[htbp]
  \centering
  \includegraphics[width=9cm]{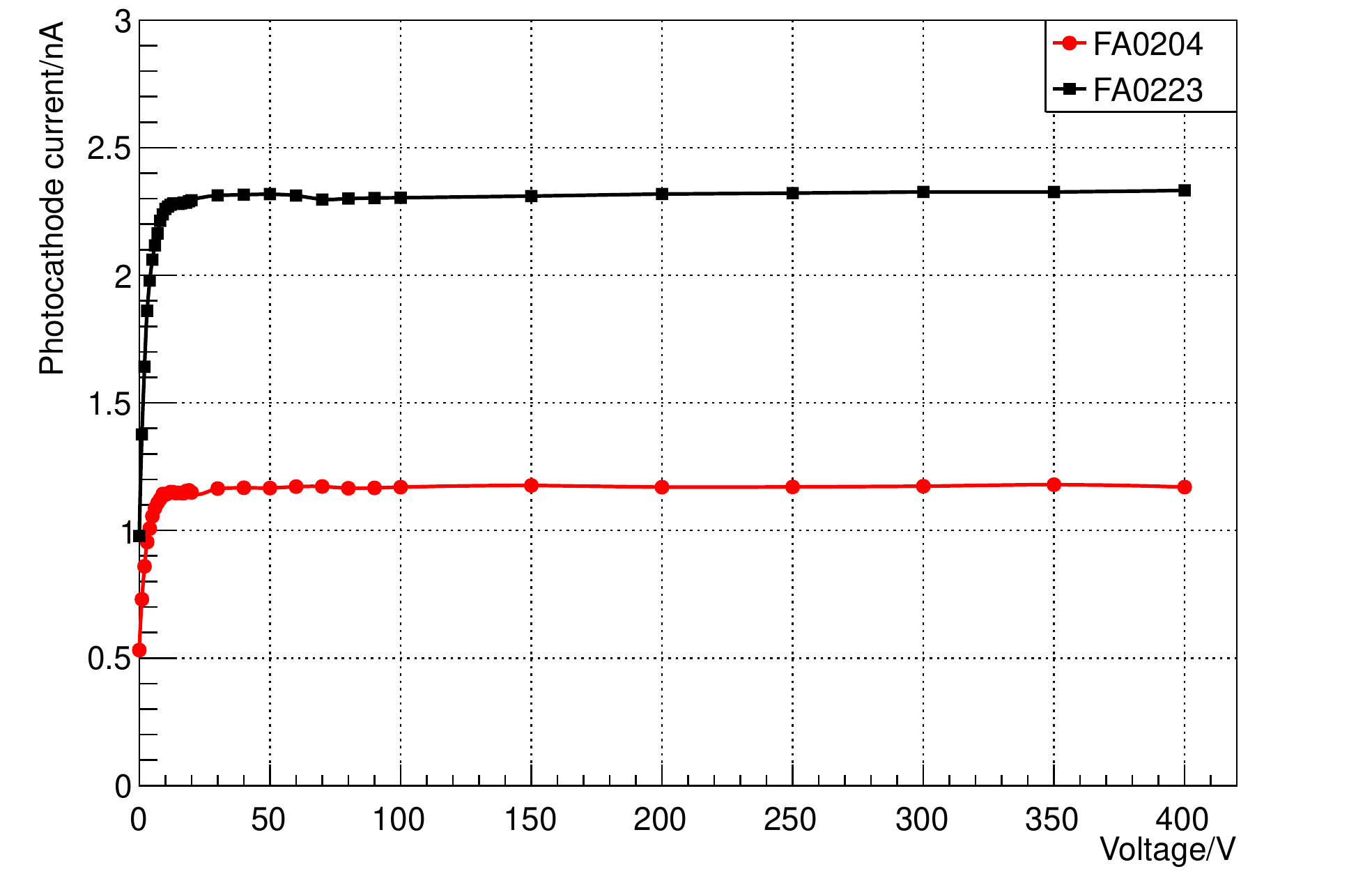}
  \caption{\label{fig:voltage_determination}The measured relationship between photocathode current and first dynode voltage at fixed light intensity. FA0204 (red dots) and FA0223(black squares) stand for the PMT serial numbers. The light intensity of the two tests is different.}
\end{figure}

It is necessary to ensure that the PMT works under appropriate voltage and incident light intensity. Based on the actual dynamic range of PMT required by the liquid argon detector, the maximum number of photoelectrons detectable when the PMT works in pulse mode is about 1000. Accordingly, if the PMT works in direct current mode, the relevant maximum magnitude of photocathode current is estimated to be about a few nanoamperes. Then, the output light intensity of the LED box is controlled by the LED supply voltage to ensure the photocathode current of PMT is under several nanoamperes. Since the wavelength of maximum response of the PMT is 420nm, 410-415nm LED is used to measure the working voltage between the photocathode and the first dynode. The measured relationship of photocathode current and the first dynode voltage is shown in the Figure.$\ref{fig:voltage_determination}$. It is found that after 50V, the photocathode current reaches a platform. Finally, 150V is selected as the working voltage of PMT in the whole test.

 A series curves of photocathode current corresponding to different light intensities are measured and shown in Figure.$\ref{fig:error}$, and the curve is fitted to obtain the slope. We consider the slope as a quantity proportional to the QE. The relative change of QE can be obtained by comparing the slope at LAr temperature and room temperature. Meanwhile, the measured data can check whether the PMT works in a linear range, which proves whether the light source has appropriate light intensity.

\section{Error discussion and data processing}

For the measurement error, the following tests are carried out.

\begin{figure}[htbp]
    \centering
    \includegraphics[width=9cm]{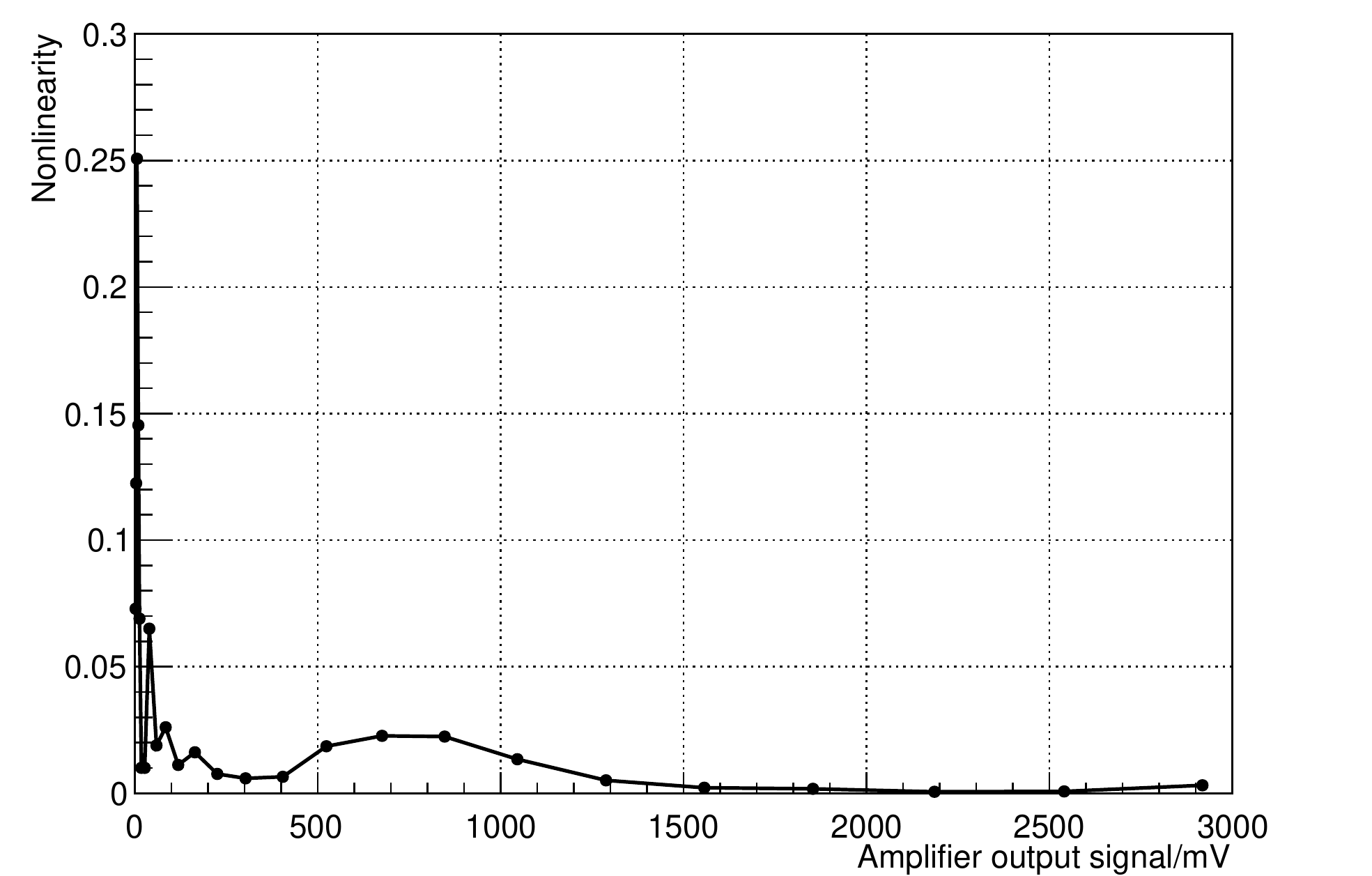}
    \caption{\label{fig:nonlinearity of amplifier} The nonlinearity of the light intensity detection system of the photosensor and amplifier.}
\end{figure}

(1)The nonlinearity of the light intensity detection system composed of the photosensor and amplifier is tested. The average peak-to-peak value $V_a$ and $V_b$ of the output signals are recorded when LED A and LED B is turned on respectively. When A and B are turned on at the same time, the average peak-to-peak value $V_c$ is recorded. The nonlinearity is defined as\cite{ZHONG}:

$$Nonlinearity = \frac{V_c- ( V_a+V_b )}{V_c}$$
Constantly changing the LED light intensity, the test results are shown in the Figure.$\ref{fig:nonlinearity of amplifier}$.

(2)The stability of the photosensor and the amplifier output signal. Under the same light intensity (trying to making the photocathode current close to 1nA), the output signal of the photosensor and amplifier is continuously recorded for about 30 min, with a total of 80 samples. The standard deviation is calculated to be 0.5$\%$.

(3)The uncertainty of the photocathode current. Under the same light intensity, the photocathode current readings of the picoammeter are continuously recorded. An average sample is obtained for every 20 numbers recorded every 20 seconds. A total of 60 samples are recorded, and the standard deviation is calculated as 0.08 $\%$.

\begin{figure}[htbp]
  \centering
  \includegraphics[width=7cm]{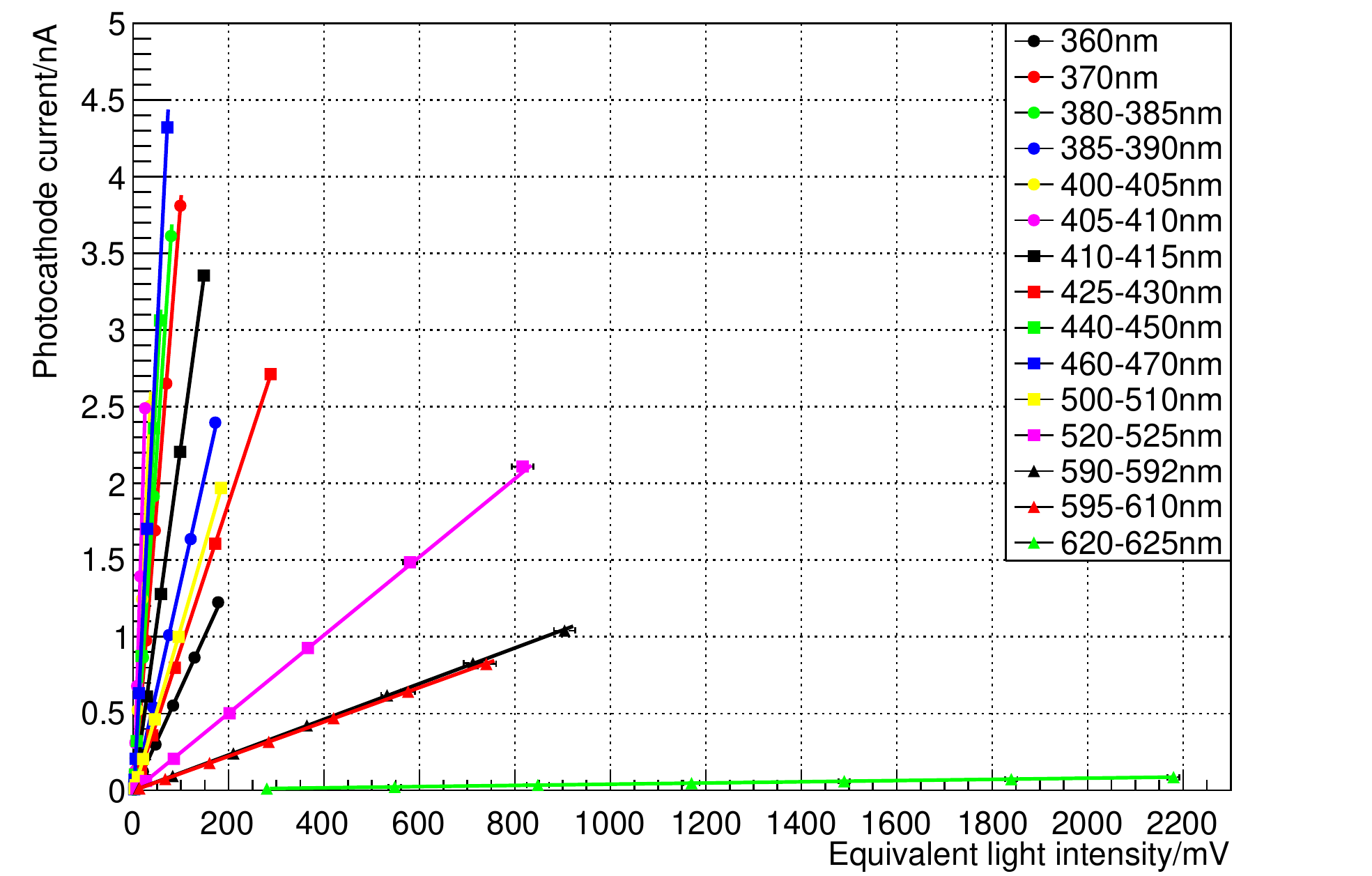}
  \qquad
  \includegraphics[width=7cm]{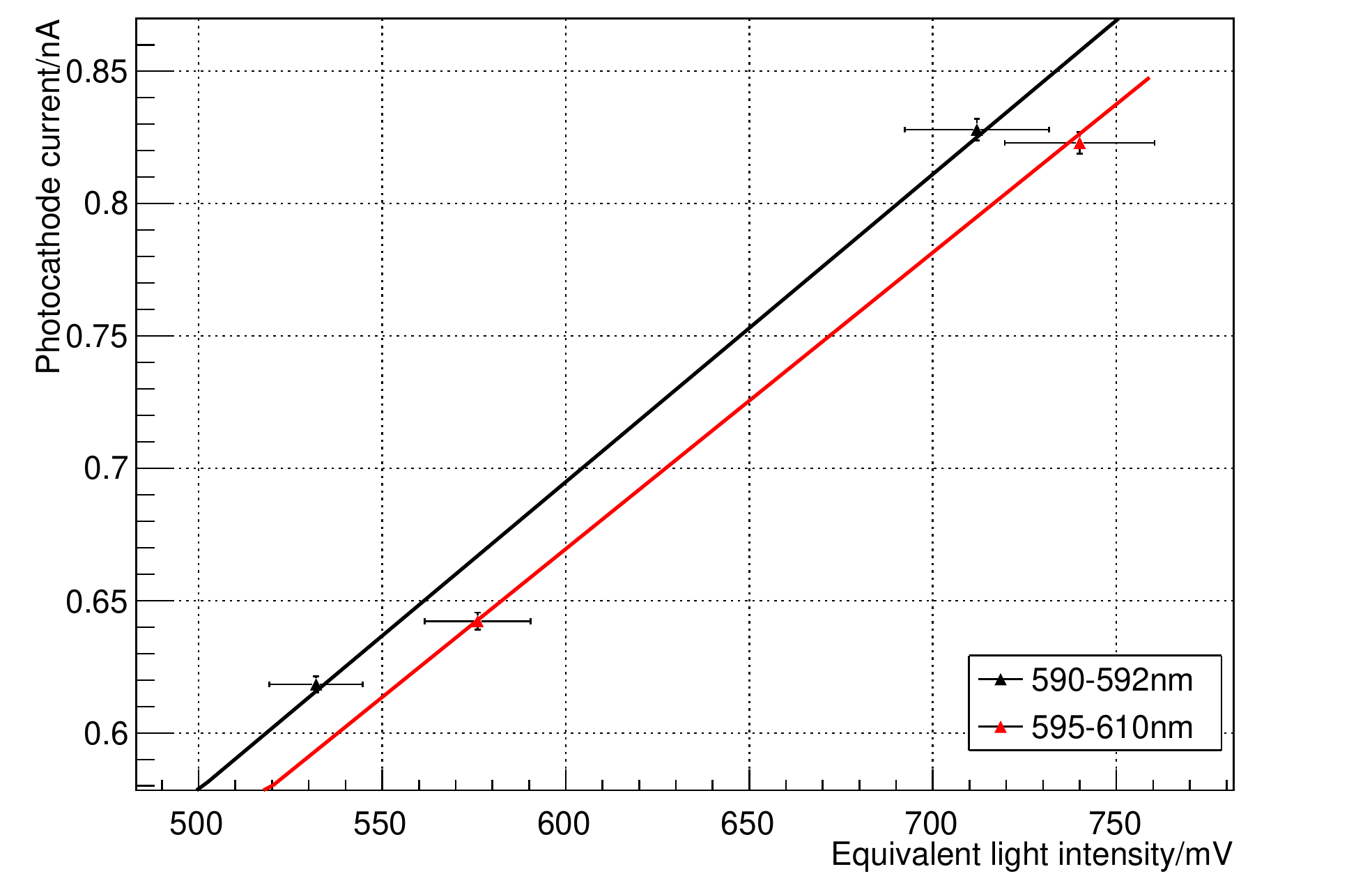}
  \caption{\label{fig:error} Left: The measured curves of photocathode current versus the equivalent light intensity at room temperature. The slope of the fitted line stands for the equivalent QE. Right: The partial enlarged figure shows the details of errors.}	
\end{figure}

As shown in the two figures in Figure.$\ref{fig:error}$, for each point on the fitting line, the X-axis is the equivalent light intensity and the Y-axis is the photocathode current. The right of Figure.$\ref{fig:error}$ is a partial enlargement of the left one to show the details of the errors. The errors of X-axis are given in two parts: the first part is the nonlinearity of the light intensity detection system. The second part is given by the stability of the detection system. Both of them give the uncertainty of the equivalent light intensity value. The errors of Y-axis include the measured uncertainty of the photocathode current and 0.4$\%$ accuracy of picoammeter from the datasheet. These points with errors are fit using straight-line function. In Figure.$\ref{fig:QE_wavelength}$, the relative QE is the ratio of equivalent QE at LAr temperature and room temperature. It is considered that the QE is independent at different temperature, and the error is given by the error transfer formula. The errors of X-axis show the wavelength ranges of each LED given by the LED datasheet in Table.$\ref{LED}$.

\section{Results and discussion}
\begin{figure}[htbp]
  \centering
  \includegraphics[width=9cm]{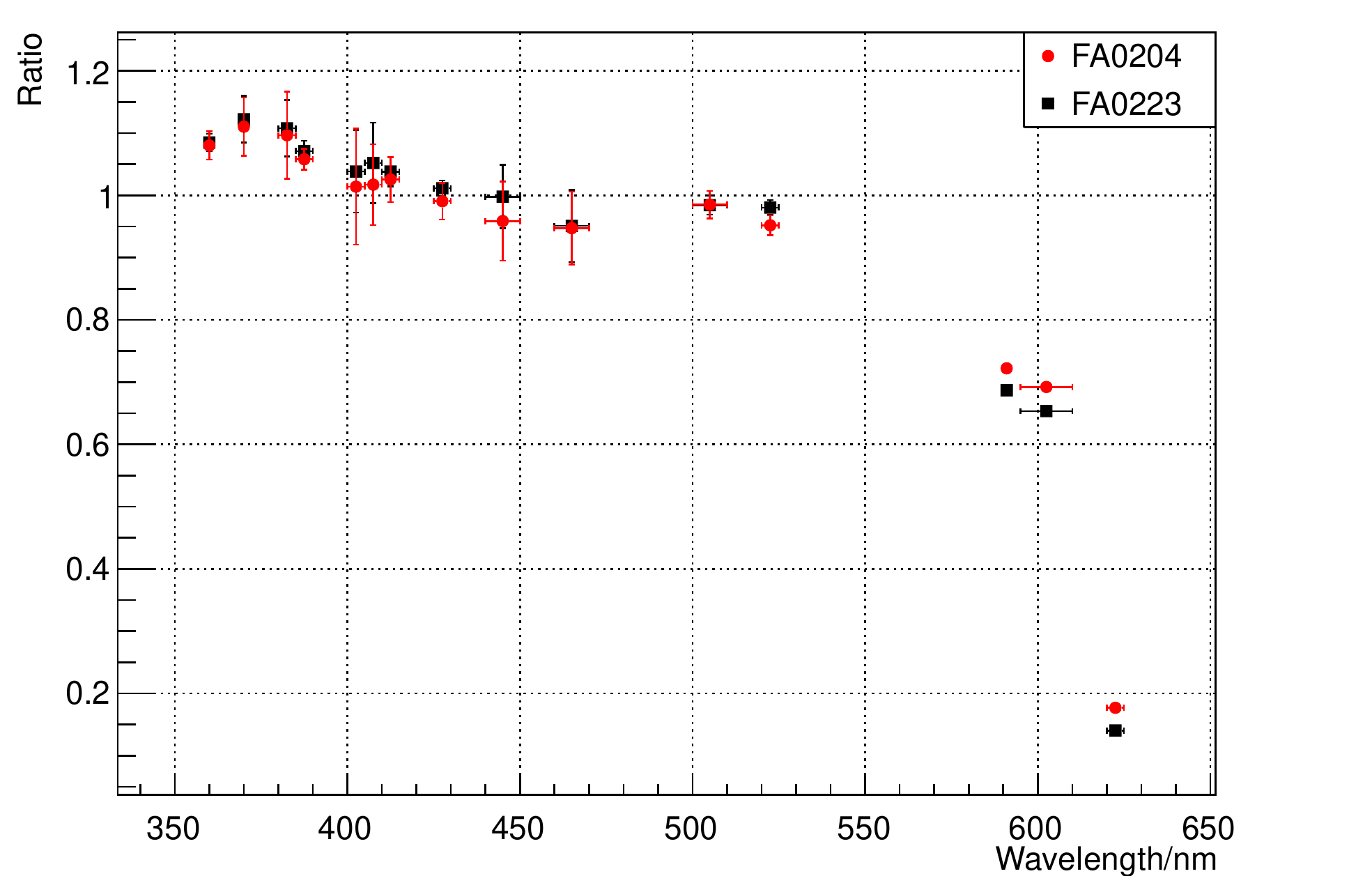}
  \caption{\label{fig:QE_wavelength} The ratio of equivalent QE at LAr temperature and room temperature versus wavelength. FA0204 (red dots) and FA0223(balck squares) stand for the PMT serial numbers. The errors of X-axis stand for the spectral ranges of each LED given by the LED datasheet.}
\end{figure}
\begin{figure}[htbp]
  \centering
  \includegraphics[width=10cm]{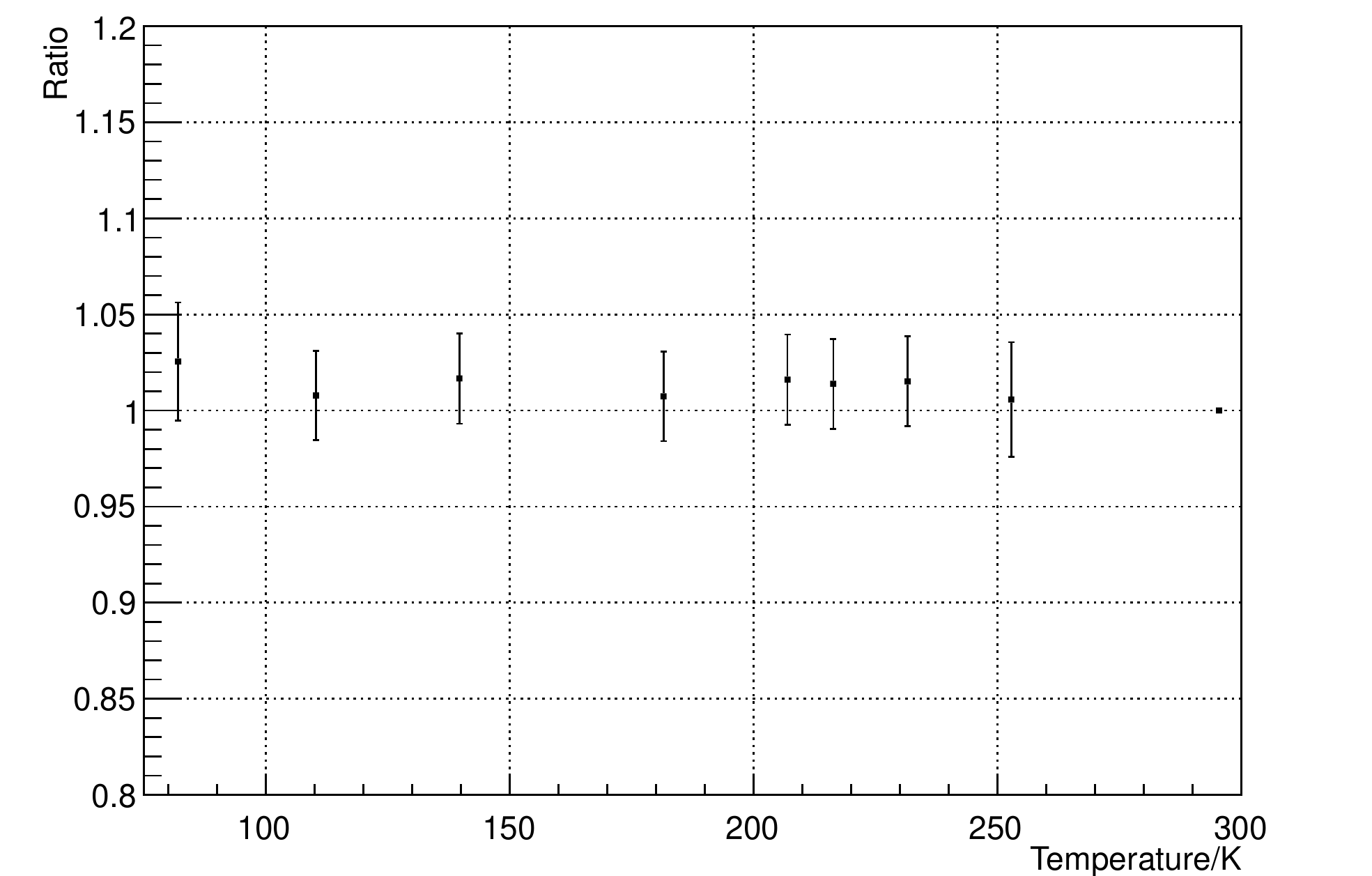}
  \caption{\label{fig:QE_temperature} The measured relative QE varying with temperature at 410-415nm. The equivalent QE at room temperature is set to 1. Temperature fluctuations during measurement are used as X-axis errors, but they are too small to be shown in the figure.}
  \end{figure}

Two PMTs are randomly selected with serial number: FA0204 and FA0223. The test results are shown in Figure.$\ref{fig:QE_wavelength}$. Around 420nm, the change of QE at LAr temperature is within about 5$\%$ compared with that at room temperature. In the shorter wavelength range, the QE increases about 10$\%$. And after 550 nm, it can be seen that the QE decreases significantly at LAr temperature. Also, the influence of different temperatures on the relative QE of one of the PMT at 410-415nm is shown in Figure.$\ref{fig:QE_temperature}$. The influence of temperature on the relative QE is less than 6$\%$.

The PMTs of R5912 series are made of an $8^{''}$ diameter borosilicate glass window with bialkali photocathode. The photocathode of PMTs with MOD in the model name is equipped the transparent Platinum under-coating that can restore the photocathode conductivity at low temperature\cite{MOD_QE}. R5912-MOD and R5912-02MOD are available with 10 and 14 dynode stages respectively. Their quantum efficiencies are 18$\%$ at 400nm\cite{MOD_QE} at room temperature. The key issue of this paper is that we have obtained the relative change of QE of the R5912-20MOD PMT at LAr temperature compared with room temperature. So the room temperature QE given by Hamamatsu still has certain reference value in practical cryogenic application. The PMTs can work normally at LAr temperature, and the performance parameter of QE can achieve the expected goal.

 The method introduced in this paper is suitable for relative QE measurement of other photoelectric devices, such as some cryogenic PMTs and SiPMs. Subsequently, we will further make some improvement aiming for measureing the absolute QE.

\acknowledgments

The study is supported by the National Key Research and Development Program of the People's Republic of China(2016YFA0400304).

The authors would like to thank Zeyuan Yu, Haiqiong Zhang for the helpful discussion.


\bibliography{references}


\end{document}